\documentclass[12pt,twocolumn,preprint]{iopart}
\usepackage{graphicx}
\usepackage{dcolumn}
\usepackage{color}
\usepackage{acronym}
\usepackage{ulem}

\newacro{AUG}[AUG]{ASDEX Upgrade}
\newacro{CXRS}[CXRS]{charge exchange recombination spectroscopy}
\newacro{BES}[BES]{beam emission spectroscopy}
\newacro{betaN}[$\beta\rm{_N}$]{normalized beta}
\newacro{btheta}[$B\rm{_{\Theta}}$]{poloidal magnetic field}

\newacro{pe}[$p\rm{_e}$]{electron pressure}
\newacro{DBP}[$D_{\omega}$]{distribution of the observed intensity}
\newacro{ne}[$n\rm{_e}$]{electron density}
\newacro{Te}[$T\rm{_e}$]{electron temperature}
\newacro{fIF}[$f\rm{_{IF}}$]{IF bandwidth}
\newacro{Trad}[$T\rm{_{rad}}$]{radiation temperature}
\newacro{Ti}[$T\rm{_i}$]{ion temperature}
\newacro{DCN}[DCN]{deuterium cyanide}
\newacro{dphi}[$\Delta \varphi\rm{_{UL}}$]{differential phase angle}
\newacro{ECE}[ECE]{electron cyclotron emission}
\newacro{ECFM}[ECFM]{electron cyclotron forward modeling}
\newacro{ECEI}[ECE-I]{ECE-imaging}
\newacro{IF}[IF]{intermediate frequency}
\newacro{inte}[$I\rm{_{imp}}$]{measured line intensity}

\newacro{nustar}[$\nu^{\star}$]{collisionality}

\newacro{zeta}[$\zeta$]{numbers of toroidal angles}

\newacro{FEM}[FEM]{finite elements}
\newacro{FILD}[FILD]{fast ion loss detector}

\newacro{ICRH}[ICRH]{ion cyclotron resonance heating}
\newacro{IDE}[IDE]{integrated data analysis equilibrium}

\newacro{ECRH}[ECRH]{electron cyclotron resonance heating}
\newacro{P_ECRH}[$P\rm{_{ECRH}}$]{ECRH power}
\newacro{OH}[OH]{ohmic heating}
\newacro{P_OH}[$P\rm{_{OH}}$]{ohmic heating power}
\newacro{PSL}[PSL]{passive stabilization loop}
\newacro{P_NET}[$P\rm{_{NET}}$]{applied net heating power}
\newacro{LIB}[LIB]{lithium beam}
\newacro{LSQ}[LSQ]{least square}
\newacro{LFS}[LFS]{low field side}
\newacro{HFS}[HFS]{high field side}
\newacro{RFA}[RFA]{resonant field amplification}
\newacro{REF-O}[REF-O]{O-mode reflectometer}
\newacro{REF-X}[REF-X]{X-mode reflectometer}
\newacro{rhotor}[$\rho\rm{_{tor}}$]{normalized toroidal flux}

\newacro{LCFS}[LCFS]{last closed flux surface}
\newacro{ELM}[ELM]{edge localized mode}
\newacro{H-mode}[H-mode]{high confinement mode}
\newacro{L-mode}[L-mode]{low confinement mode}

\newacro{PCS}[PCS]{plasma control system}

\newacro{WMHD}[$W\rm{_{MHD}}$]{plasma energy}

\newacro{MP}[MP]{magnetic perturbation}
\newacro{IDA}[IDA]{integrated data analysis}
\newacro{L-mode}[L-mode]{low confinement mode}
\newacro{NBI}[NBI]{neutral beam injection}
\newacro{SOL}[SOL]{scrape off layer}
\newacro{LOS}[LOS]{lines of sight}
\newacro{LC}[$L\rm{_C}$]{Connection length}
\newacro{MHD}[MHD]{magnetohydrodynamics}

\newacro{TS}[TS]{Thomson scattering}

\begin{document}

\title[\small 3D boundary displacement due to stable ideal kink modes excited by external n=2 \acsp{MP} ]{Three dimensional boundary displacement due to stable ideal kink modes excited by external n=2 \aclp{MP}}

\author{M. Willensdorfer$^1$\footnote{matthias.willensdorfer@ipp.mpg.de}, E.~Strumberger$^1$, W.~Suttrop$^1$, M.~Dunne$^1$, R.~Fischer$^1$,
G.~Birkenmeier$^{1,2}$, D.~Brida$^{1,2}$, M.~Cavedon$^{1}$, S.~S.~Denk$^{1,2}$, V.~Igochine$^1$,  L.~Giannone$^1$, A.~Kirk$^3$, J.~Kirschner$^{1,4}$, A.~Medvedeva$^{1,2,5,6}$, T.~Odstr\v{c}il$^{1,2}$, D.~A.~Ryan$^{3,7}$, the ASDEX Upgrade Team$^1$ and the EUROfusion MST1 Team$^*$}

\address{ $^1$ Max Planck Institute for Plasma Physics, 85748 Garching, Germany \\
$^2$ Physik-Department E28, Technische Universit\"at M\"unchen, 85748 Garching, Germany \\
$^3$ CCFE, Culham Science Centre, Abingdon, Oxon, OX14 3DB, UK \\
$^4$ Institute of Applied Physics, TU Wien, Fusion@\"OAW, Austria \\
$^5$ Institut Jean Lamour UMR 7198 CNRS, UniversitŽ de Lorraine, F-54000, Nancy, France \\
$^6$ CEA, IRFM, F-13108, St-Paul-Lez-Durance, France \\
$^7$York Plasma Institute, Department of Physics, University of York, Heslington, York,YO10 5DQ, UK \\

$^*$ See the author list of ÒOverview of progress in European Medium Sized
Tokamaks towards an integrated plasma-edge/wall solutionÓ by H. Meyer
et al, to be published in Nuclear Fusion Special issue: Overview and Summary
Reports from the 26th Fusion Energy Conference (Kyoto, Japan, 17-22 October 2016)    \vspace{-8pt}
 }


\begin{abstract}

In low-\ac{nustar} scenarios exhibiting mitigation of \acp{ELM}, stable ideal kink modes at the edge are excited by externally applied \ac{MP}-fields. At ASDEX Upgrade these modes can cause three-dimensional (3D) boundary displacements  up to  the centimeter range. These displacements have been measured using toroidally localized high resolution diagnostics and rigidly rotating $n=2$ \ac{MP}-fields with various applied poloidal mode spectra. These measurements are compared to non-linear 3D ideal \ac{MHD} equilibria calculated by VMEC. Comprehensive comparisons have been conducted, which consider for instance plasma movements due to the position control system, attenuation due to internal conductors and changes in the edge pressure profiles.

VMEC accurately reproduces the amplitude of the displacement and its dependencies on the applied poloidal mode spectra. Quantitative agreement is found around the \ac{LFS} midplane.  The response at the plasma top is qualitatively compared. The measured and predicted displacements at the plasma top  maximize when the applied spectra is optimized for \ac{ELM}-mitigation.
The predictions from the vacuum modeling generally fails to describe the displacement at the \ac{LFS} midplane as well as at the plasma top. When the applied mode spectra is set to maximize the displacement, VMEC and the measurements clearly surpass the predictions from the vacuum modeling by a factor of four.
Minor disagreements between VMEC and the measurements are discussed.
This study underlines the importance of the stable ideal kink modes at the edge for the 3D boundary displacement in scenarios relevant for \ac{ELM}-mitigation.

\end{abstract}

\maketitle

\acresetall

\section{Introduction}

Externally applied \acp{MP} can be used to mitigate and to suppress \acp{ELM} in \ac{H-mode}~\cite{Evans:2004}. At low \acl{nustar} ($\acs{nustar}<0.3$), \ac{ELM} mitigation and suppression are accompanied with a loss of confinement primarily resulting in the loss of density, the so-called density 'pump-out'. Recent studies at ASDEX Upgrade~\cite{Kirk:2015, Suttrop:2017}, DIII-D~\cite{Paz-Soldan:2015} and MAST~\cite{Kirk:2015} have shown that both the best \ac{ELM} mitigation as well as suppression are achieved by an externally applied \ac{MP}-field when its poloidal mode spectrum excites modes at the edge which are most amplified by the plasma. 
According to \ac{MHD} calculations~\cite{Liu:2011}, these modes are stable ideal kink modes, which are driven by the \ac{H-mode} edge pressure gradient and/or the associated bootstrap current~\cite{Paz-Soldan:2016}. 
Because of the amplification by the plasma, the resulting \acp{MP} at the plasma boundary can be even larger than expected solely from the externally applied \acp{MP}~\cite{Moyer:2012, King:2015}. 
Moreover, these stable kink modes cause a 3D displacement of the plasma boundary, which is clamped to the applied MP field.  \ac{MHD} codes, like IPEC~\cite{Park:2007}, JOREK~\cite{Orain:2017}, MARS-F~\cite{Liu:2000}, M3D-C1~\cite{Ferraro:2013}, VMEC~\cite{Hirshman:1983,Strumberger:2014} are able to calculate this deformation for various plasma scenarios and coil configurations. These \ac{MHD} codes predict that the X-point displacement, also referred as \ac{HFS} response~\cite{Paz-Soldan:2016} or peeling response, maximizes when the applied poloidal mode spectrum is optimized for \ac{ELM} mitigation or suppression. It is therefore assumed that the X-point displacement influences the \ac{ELM} stability.

The characterization and  prediction of the non-axisymmetric boundary deformation is important because such 3D geometry can influence the \ac{ELM} stability~\cite{Chapman:2013}, turbulent transport~\cite{Bird:2013} and the coupling of the \ac{ICRH}~\cite{Bobkov:2014}. The 3D boundary distortion from external \acp{MP} has been extensively studied in various machines like ASDEX Upgrade~\cite{Fischer:2012, Fuchs:2014}, DIII-D~\cite{Lanctot:2011, Moyer:2012, Ferraro:2013}, MAST~\cite{Chapman:2012}, JET~\cite{Chapman:2007,Yadykin:2015} and has been reviewed in Ref.~\cite{Chapman:2014a}. The main conclusion of Ref.~\cite{Chapman:2014a} is that the measured displacement of the \ac{LFS} midplane boundary depends approximately linearly on the applied resonant field predicted by vacuum field modeling \cite{Chapman:2014a}. But it was also observed that in some cases the vacuum modeling clearly underestimates the displacement due to stable ideal kink modes.

In this paper, we demonstrate that in a scenario which exhibits \ac{ELM} mitigation at low \acs{nustar} 
stable ideal kink modes dominate the boundary displacement. When the applied poloidal mode spectrum is optimized to excite the edge kink mode, the displacement is about four times larger than the  expectation from the vacuum field modeling. Thus, predictions by vacuum field modeling are not a good approximation. This is similar to one case studied in Ref.~\cite{Moyer:2012, Orlov:2014}. We extended the analysis of  Ref.~\cite{Willensdorfer:2016} and present comprehensive studies  of the 3D boundary displacement at ASDEX Upgrade using rigidly rotating \ac{MP}-fields with toroidal mode number $n=2$ and toroidally localized diagnostics. The analysis methods have been further improved including the effects from the plasma position control system, the attenuation of the \ac{MP}-field from internal conductors and the applied poloidal mode spectrum. This allows us to achieve a new level of accuracy and to perform detailed analyses of the local plasma response via the displacement.
We further characterize the dependence of the 3D displacement on the applied poloidal mode spectra by varying the \ac{dphi}~\cite{Suttrop:2011b}, which is the toroidal phase of the \ac{MP}-field from the upper coil set $\varphi\rm{_U}$ subtracted by the lower one $\varphi\rm{_L}$, $\ac{dphi}=\varphi\rm{_U}-\varphi\rm{_L}$. These measurements are compared to the results of the non-linear ideal \ac{MHD} equilibrium code VMEC, which has also been employed at other devices~\cite{Chapman:2012,Lazerson:2015, Wingen:2017, Koliner:2016}. It is demonstrated that VMEC can predict quantitatively the displacement. 

This paper is organized as follows. Section \ref{sec:setup} describes the experimental configuration and measuring principles. The modeling is described in section~\ref{sec:modelling}. In section \ref{sec:ELMbehavior}, we test the plasma response via the \ac{ELM} behavior. This is then compared  to displacement measurements and calculations around the plasma top in section~ \ref{sec:plasmatop}. The comparison between measurements and modeling of the displacement around the \ac{LFS} midplane is shown in section \ref{sec:midplane}. This paper concludes with section \ref{sec:conclusion}. The sensitivity studies regarding the grid resolution in VMEC are shown in the Appendix.
%

\section{Experimental configuration}
\label{sec:setup} 

\subsection{Discharge configuration with rigid rotations}
\label{svec:discharge} 

The present configuration is very similar to the one studied in Ref.~\cite{Willensdorfer:2016}. The experiments have a toroidal field $B\rm{_T}$ of $-2.5\ \rm{T}$, low triangularity (lower $\delta\rm{_l} =0.52$ and upper $\delta\rm{_u}=0.119$) and a plasma current of $800\ \rm{kA}$ resulting in a safety factor of $q_{95}\approx-5.2$. In addition to the ohmic heating of about $300\;\rm{kW}$, the external heating power in the discharges presented here amounts to around $7\;\rm{MW}$ from \ac{NBI} and $2\;\rm{MW}$ from centrally deposited \ac{ECRH}. 

\begin{figure}[ht]
   \centering
 \includegraphics[width=0.48\textwidth]{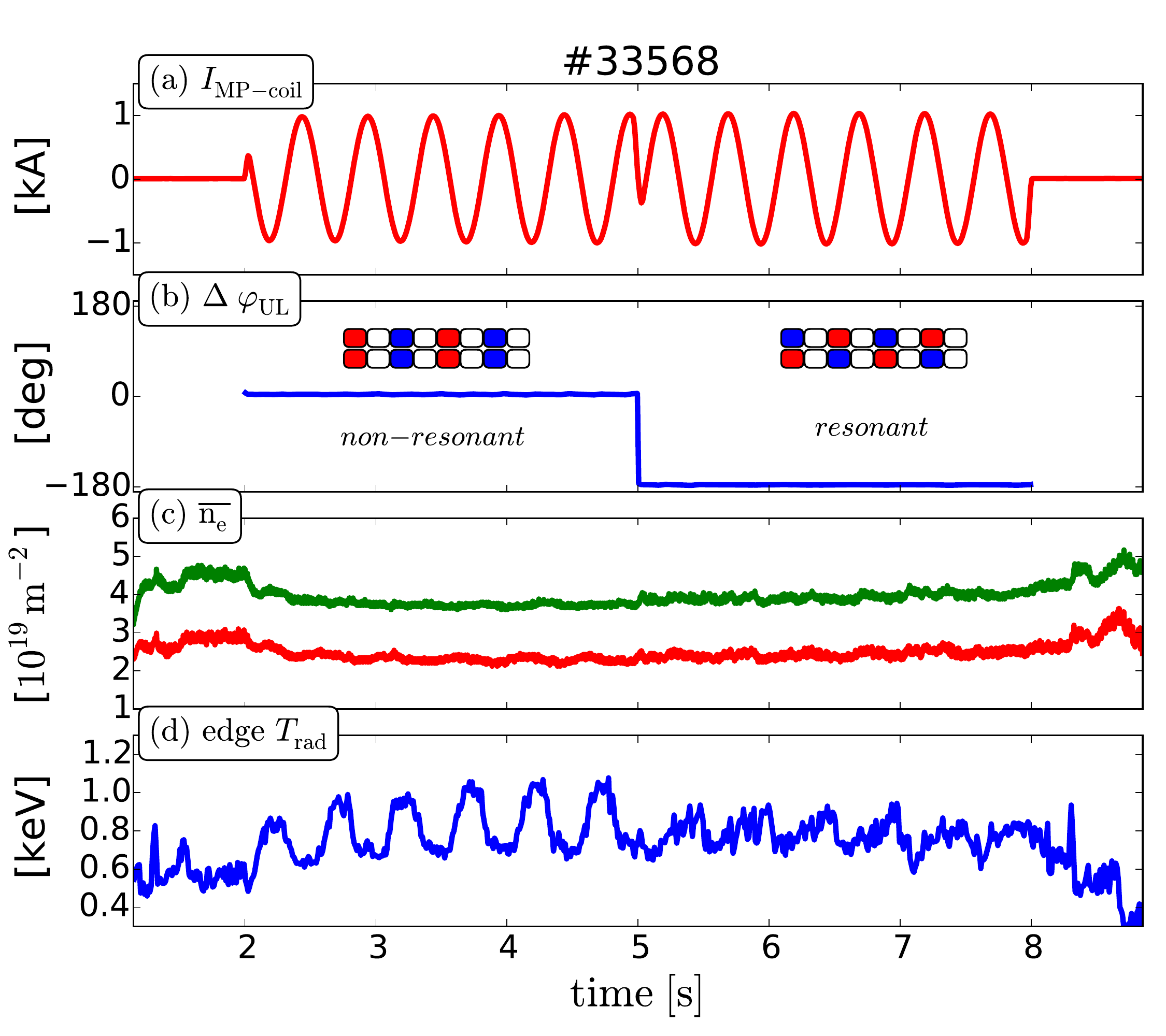} 
 \caption{Overview of a typical discharge with rigid rotation: (a) Power supply current of one \ac{MP}-coil illustrating the timing of the rigid rotation with $2\ \rm{Hz}$, (b) the differential phase angle of two rotation phases employing vacuum \textit{non-resonant} and \textit{resonant} configuration, (c) line integrated density of a core (green) and edge chord (red) and (d)  \ac{Trad}($\approx\ac{Te}$) from \ac{ECE} in the pedestal around the \ac{LFS} midplane. The modulation amplitude measured by \ac{ECE} depends clearly on \ac{dphi}. }
\label{fig:DischargeOverview}
\end{figure}

The applied $n=2$ \ac{MP}-fields are produced by 16 saddle coils with 8 coils in each row (see Ref.~\cite{Suttrop:2011b}).
To measure the displacement using toroidally localized diagnostics, we rotate the applied \ac{MP}-field rigidly. Figure \ref{fig:DischargeOverview} shows time traces of a typical discharge. To indicate the timing of the rigid rotation, the top frame shows the supplied current of one \ac{MP}-coil. To test the plasma response for different applied poloidal mode spectra, we varied \ac{dphi} in-between discharges and, in some cases,  during the discharge. In the illustrated discharge, the external \ac{MP}-field rotates rigidly with $2\;\rm{Hz}$ at two different values for \ac{dphi} for 3 seconds each. First, \ac{dphi} of $\approx 0^{\circ}$  was applied,  which is close to the maximum mis-alignment of the external \ac{MP}-field with respect to the equilibrium field in the pedestal. Therefore, we refer to this configuration as (vacuum) \textit{non-resonant} in figure\ \ref{fig:DischargeOverview}(b). Then, at 5 seconds, we set \ac{dphi} to $\approx \pm180^{\circ}$, which is the optimum field-alignment and therefore, labelled as (vacuum) \textit{resonant}. 
During both phases a moderate degree of density 'pump-out' ($10-20\%$) is observed as shown in the measured line integrated densities using the edge and core chord (figure~\ref{fig:DischargeOverview}(c)).
The time trace of one edge \ac{ECE} channel exhibits a clear modulation in the measured \ac{Trad}, which is caused by the radial displacement (figure~\ref{fig:DischargeOverview}(d)). This channel is optically thick, so we can assume that \ac{Te} can be approximated by \ac{Trad}.
Furthermore, the amplitude clearly changes with the applied \ac{dphi}.


\subsection{Set of discharges with rigid rotations}

\begin{table}
\begin{tabular}{c|c|c|c|c|c}
shot & set $\Delta \varphi \rm{_{UL}}$ [$^\circ$]  & f [Hz] &  $<\beta_N>$ &  available diagnostics \\ \hline
\textcolor{red}{33118} &\textcolor{red}{-90}  & 3 & \textcolor{red}{1.82} & ECE, CXRS, LIB, REF-X\\ 
\textcolor{blue}{33345} & \textcolor{blue}{90} , -90 & 2 & \textcolor{blue}{2.18}, 1.81 & ECE, CXRS, LIB, REF-X\\
33346 & 130, 50 & 2 & 2.27, 2.09 & ECE, CXRS, REF-X \\
33568 & 0, $\pm180$ & 2 & 1.83, 1.96 &  ECE \\
\textcolor[rgb]{0,0.65,0}{33569} & \textcolor[rgb]{0,0.65,0}{-50} , -130 & 2 & \textcolor[rgb]{0,0.65,0}{2.01}, 2.03 & ECE, CXRS, LIB, REF-X\\
33570 &  -100 & 2 & 2.0 &  ECE, CXRS, REF-X  \\
\end{tabular}
 \caption{ Overview of analyzed discharges. The experimental periods used as input for the \ac{MHD} modeling are colored.  $<\beta_N>$ is the normalized beta averaged over the analyzed rotation period. }
\label{tab:discharges}
\end{table}

To systematically study the plasma response during \ac{ELM} mitigation, we applied rigidly rotating \ac{MP}-fields  with various \acp{dphi}. The resulting change of the applied poloidal mode spectra allows us to investigate its impact on the plasma response and the non-axisymmetric boundary displacement. In all discharges, the same plasma shape, heating power and gas fuelling rate was configured. Only in discharge $\#33569$, one gyrotron tripped prior to the rotation phase resulting in $500\ \rm{kW}$ less \ac{ECRH} power. 
The set of discharges with the different phases of \ac{dphi} is listed in Table~\ref{tab:discharges}. Because the set  \ac{dphi} also  influences the  density 'pump-out'~\cite{Paz-Soldan:2015}, the  density and hence, \ac{betaN} vary by around $20\%$.




\subsection{Displacement measurements around the midplane}
\label{midplane_displacement}
\begin{figure}[ht]
   \centering
 \includegraphics[width=1.0\textwidth]{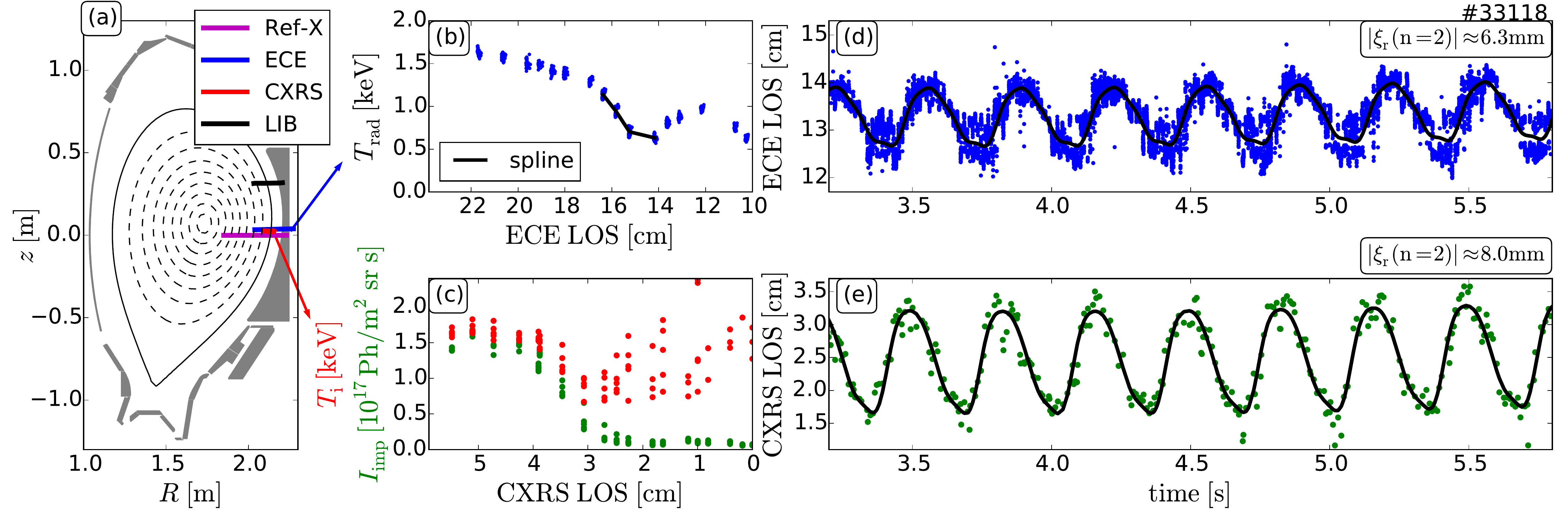} 
 \caption{Measuring  $\xi_r$ around the \ac{LFS} midplane. (a) poloidal cross-section showing \ac{LOS} of \ac{REF-X}, \ac{ECE}, \ac{CXRS} and \ac{LIB}, (b) \ac{Trad} profile from \ac{ECE} and spline (line) to determine profile shift, (c) \ac{Ti} profile in red and \ac{inte} in green from \ac{CXRS}. (d) and (e) show time traces of the separatrix position from \ac{ECE} and \ac{CXRS} as well as their corresponding sine fits (line), respectively.} 
\label{fig:evalDisplacement}
\end{figure}

To measure the radial displacement, we use the high resolution profile diagnostics around the \ac{LFS} midplane. Figure~\ref{fig:evalDisplacement}(a) shows the set of used diagnostics consisting of profile \ac{ECE}~\cite{Rathgeber:2013, Willensdorfer:2016}, \ac{LIB}~\cite{Willensdorfer:2012, Willensdorfer:2014}, edge \ac{CXRS}~\cite{Viezzer:2012} and \ac{REF-X}~\cite{Medvedeva:2016}. As some diagnostic data were not available for all discharges, the last column of Table~\ref{tab:discharges} lists the availability of the various profile diagnostics.

 To determine the displacement around the midplane, we track movements in the profile diagnostics during the rigid rotation using only pre \ac{ELM} data points ($60-90\%$ of the \ac{ELM} cycle).
In the case of \ac{ne} profile measurements, the procedure is straight forward. We determine the  density~\cite{Willensdorfer:2016} at the separatrix before the \ac{MP}-phase and track the position of this values along the diagnostic \ac{LOS} during the rigid rotation. In our case, it is $1.2\cdot 10^{19}\ \rm{m^{-3}}$ and is used for all \ac{ne} profile diagnostics and for all analyzed phases. Small variations of this density value do not change the outcome of the analysis because of the steep density gradients in the pedestal. 

For \ac{CXRS} measurements a similar method is used. But instead of using the \ac{Ti} or the rotation profiles, it is more advantageous to use the \ac{inte}, which is Boron 5+   ($B_{5+}$) in this case.  \ac{Ti} and rotation profiles are not reliable in the \ac{SOL}, because of a low \ac{inte}. They usually exhibit large uncertainties and a large scatter in the \ac{SOL} (see figure~\ref{fig:evalDisplacement}(c)). Because of the low beam attenuation at the edge, \ac{inte} is approximately proportional to the impurity density around the plasma boundary.  Therefore, the \ac{inte}-profile increases monotonically  from the \ac{SOL} towards the pedestal top (figure~\ref{fig:evalDisplacement}(c)). This allows us to use the same procedure for \ac{CXRS} as for the \ac{ne} profile measurements. A value of $0.5\cdot 10^{17}\ \rm{Ph/m^2sr\ s}$ at the separatrix is determined prior to the \ac{MP}-phase and is used for all cases. 

\ac{ECE} measurements require a different approach due to the non-monotonic behavior of the \ac{Trad} profile from the \ac{ECE} diagnostic at the edge known as the 'shine-through' effect~\cite{Rathgeber:2013}. 
To obtain the plasma displacement, first, the \ac{Trad} data from the steep gradient region is fitted using a spline at the beginning of each rigid rotation phase~\cite{Willensdorfer:2016}. Then, this spline is only varied by a radial shift until the \ac{LSQ} is minimized (see also Ref.~\cite{Willensdorfer:2016}). This is done for every pre-\ac{ELM} time point throughout the analyzed time window.  

\subsection{Estimation of the axisymmetric contribution due to the plasma position control}
\label{sec:pcs}

\begin{figure*}[ht]
   \centering
 \includegraphics[width=1.0\textwidth]{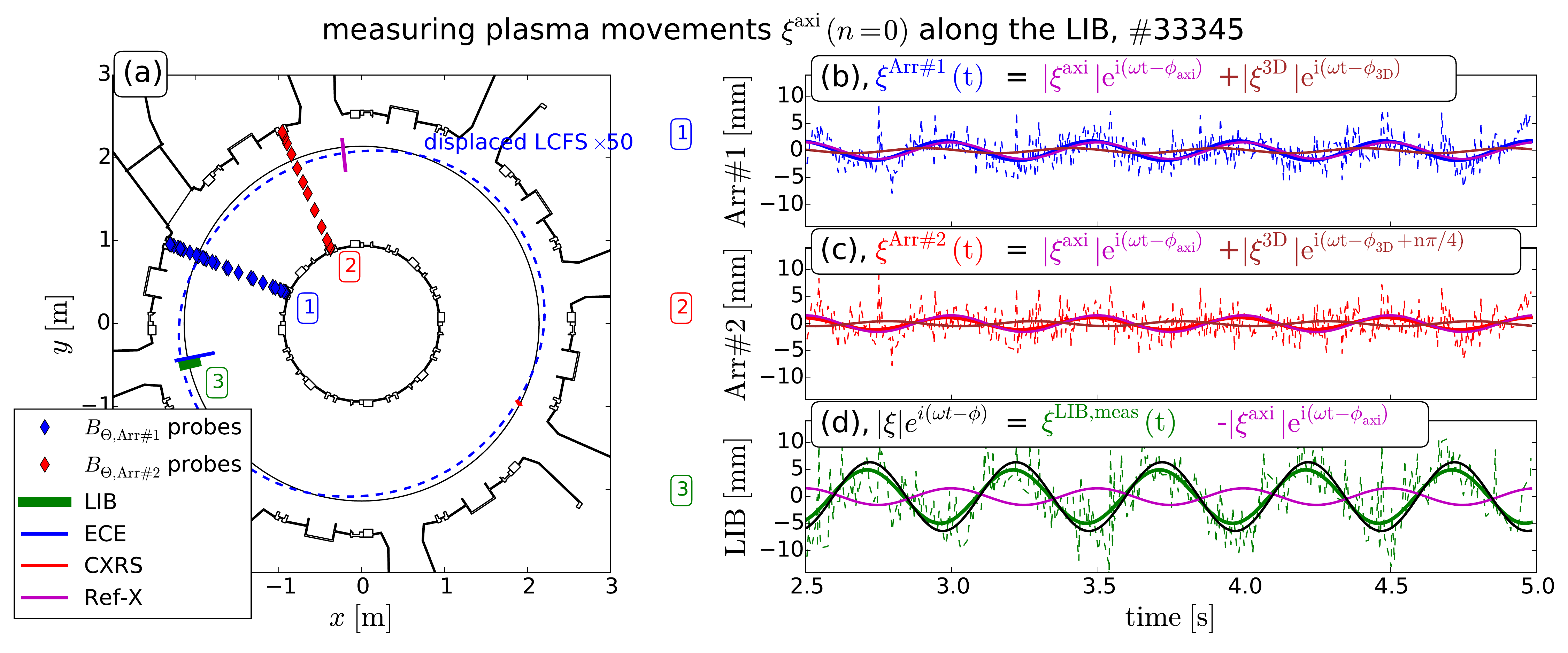} 
 \caption{Measuring $n=0$ plasma movements at the \ac{LOS} of the \ac{LIB} diagnostic. (a) top view of the experimental configuration showing the various diagnostics and two \ac{btheta} arrays.  Array 1 ($Arr\#1$, blue diamonds), which is used for the \ac{PCS} and  Array 2 ($Arr\#2$, red diamonds). Solid black and dashed blue line indicate the outer \ac{LCFS} and the displaced VMEC boundary amplified by a factor of 50, respectively. Separatrix movements at the \ac{LIB} using equilibrium reconstructions with (b) $Arr\#1$ (blue) and (c) $Arr\#2$ (red). The determined axisymmetric $\xi\rm{^{axi}}$ and non-axisymmetric contributions $\xi\rm{^{3D}}$  are purple and brown, respectively. (d) Measured displacement using \ac{LIB} (green) and determined displacement (black). In general, effects from \ac{PCS} are small.} 
\label{fig:plasmaPosition}
\end{figure*}

To guarantee a stable plasma operation during these \ac{MP}-field rotation experiments, the plasma position at the outer midplane is feedback-controlled. 
The \ac{PCS} assumes an axisymmetric equilibrium during the rigidly rotating \ac{MP}-field. Because the reconstruction of the actual radial plasma position is based on \ac{btheta} measurements localized at one toroidal position (blue diamonds in \ref{fig:plasmaPosition}(a)), it can induce additional sinusoidal axisymmetric $n=0$ movements of the plasma for two reasons. First, the \ac{btheta} probes pick up stray-fields from the \ac{MP}-coils and the field generated by the plasma response to the \ac{MP}, which are not accounted for in the realtime equilibrium regression used for the \ac{PCS}. Second, the  control system tries to counteract the rotating 3D displacement (see Ref.~\cite{Chapman:2014b}). 

These sinusoidal $n=0$ movements have the same frequency as the rotating displacement and can distort the displacement measurements in amplitude and phase. 
To interpret the measurements correctly, it is therefore necessary to quantify this contribution from the $n=0$ movements. 
This can be done by reconstructing the axisymmetric equilibrium  throughout the rigid rotation at two toroidal positions  using  two toroidally separated \ac{btheta} arrays (blue and red diamonds in figure \ref{fig:plasmaPosition}(a)). It allows us to disentangle the  $n=0$ movements from distortions in the axisymmetric equilibrium reconstruction due to the non-axisymmetric effects like the 3D displacement and the 'pick-up' in the magnetic probes. 
The main idea is that during the rigid rotation the non-axisymmetric contributions  appear in both equilibrium reconstructions with  a preset phase difference ($n\pi/4$) depending on the toroidal separation of the two probe arrays and the applied  toroidal mode number $n$, whereas the axisymmetric $n=0$ contributions appear simultaneously. This enables us to quantify the $n=0$ movements  along various \ac{LOS} of used diagnostics and subtract it from the measured $n=0$ displacement of the \ac{LCFS}.

One example of this procedure is shown in Figure \ref{fig:plasmaPosition} (discharge $\#33345$, $\ac{dphi}\approx90^\circ$). This experiment exhibits the largest plasma movements ($\pm1.5\ \rm{mm}$) within the set of discharges (Table~\ref{tab:discharges}). Figure \ref{fig:plasmaPosition}(b) and (c) show time traces of the separatrix movements along the \ac{LIB}  \ac{LOS} from the \ac{IDE}~\cite{Fischer:2016} data from the array 1 (blue) and array 2 (red), respectively. The determined axisymmetric $n=0$ contribution caused by the control system ($\xi\rm{^{axi}}$) and non-axisymmetric one ($\xi\rm{^{3D}}$) are shown in purple and brown, respectively. The extracted axisymmetric contribution $\xi\rm{^{axi}}$  is then subtracted from the \ac{LIB} measurement ($\xi\rm{^{LIB}}$, green) to determine the actual displacement ($\xi$, black) along the \ac{LIB}, which is shown in figure~\ref{fig:plasmaPosition}(d).
Please note that the \ac{LIB} is above the midplane (see figure~\ref{fig:evalDisplacement}(a)). 
This procedure is applied to each rigid rotation phase and for each used diagnostics of table~\ref{tab:discharges}. In all experiments, the observed $n=0$ movements of the plasma are relatively small ($0-1.5\ \rm{mm}$) with respect to the measured radial displacement ($2-8\ \rm{mm}$).  From these numbers, we can already conclude that the \ac{PCS} in ASDEX Upgrade, which uses only magnetic measurements, is not fully counteracting the 3D boundary displacement~\cite{Chapman:2014b}. Otherwise, the $n=0$ movements would have the same magnitude as the displacements. Detailed analysis of the behavior of the \ac{PCS} and the cause of this $n=0$ movements are beyond the scope of this paper and will be published elsewhere.

\subsection{Plasma top diagnostics for \ac{HFS} response}

As already mentioned in the introduction, the displacement around the X-point and the \ac{HFS} response, are thought to be  important for \ac{ELM} mitigation. 
However, \ac{MHD} codes with spectral representation exclude the X-point. VMEC calculations done for ASDEX Upgrade plasmas show the lowest displacement amplitude at the X-point. The X-Point region is difficult to diagnose and requires sophisticated plasma response measurements~\cite{Shafer:2014}. Instead, we use the displacement around the plasma top to characterize this \ac{HFS} response. It exhibits the same dependence on \ac{dphi} as  the X-point and the \ac{HFS} midplane (see Ref.~\cite{Liu:2016b}).

To probe the response at the plasma top, we use one soft X-ray channel with  a $75\ \mu \rm{m}$ filter~\cite{Igochine:2010}. 
The \ac{LOS} is exactly tangential to the axisymmetric flux surfaces (see figure \ref{fig:SXRplasma}(a)). To evaluate the displacement, we use again only pre-\ac{ELM} data and fit the time trace of the emissivity using the same sine function as in section \ref{fig:evalDisplacement}. The relative amplitude of the emissivity is then compared to the local displacement calculated by VMEC. Therefore, it is only possible to make a qualitative comparison.
This analysis is relatively simple and does not allow us to account for the effects from the \ac{PCS}, which were shown to be small in section \ref{sec:pcs}. 


\section{Modeling of the displacement}
\label{sec:modelling}

\subsection{Screening of transient \acp{MP} due to image currents in passive conductors}
\label{sec:PSLresponse}

ASDEX Upgrade has a \ac{PSL} to reduce the growth rate of vertical instabilities. It is a copper conductor onto which the \ac{MP}-coils are mounted. Thus, local image currents in the \ac{PSL} can attenuate and delay  transient \ac{MP}-fields at the plasma boundary depending on their frequency. To quantify the attenuation and the phase delay~\cite{Suttrop:2016a}, \ac{FEM} calculations have been employed. According to these calculations, the \ac{MP}-field amplitudes for a $2\ \rm{Hz}$ rotation are reduced to $62.1\%$ and $68.7\%$ for the upper and lower coil set, respectively, whereas for $3\;\rm{Hz}$,  they are reduced to  $56.3\%$ and $64.5\%$. The variations between the upper and lower coil set arises from slightly different positions with respect to the \ac{PSL}. 
In a rigid rotation the different \ac{PSL} responses for the upper and lower coils have  also a small effect on the differential phase \ac{dphi}, which changes by around  $-4^\circ$ for $3\;\rm{Hz}$ and even lower for $2\;\rm{Hz}$. To account for this attenuation in the modeling, we simply applied the response function from the \ac{FEM} calculations to the power supply current of the \ac{MP}-coils. 
The result is an 'effective' coil current, which is used as an input for the modeling. This approach is legitimate, since the distance between \ac{MP}-coils and \ac{PSL} is much shorter than the one between \ac{MP}-coils and the plasma.

\subsection{Input Equilibria}
\label{sec:equilibria} 

\begin{figure}[ht]
 \centering
 \includegraphics[width=0.5\textwidth]{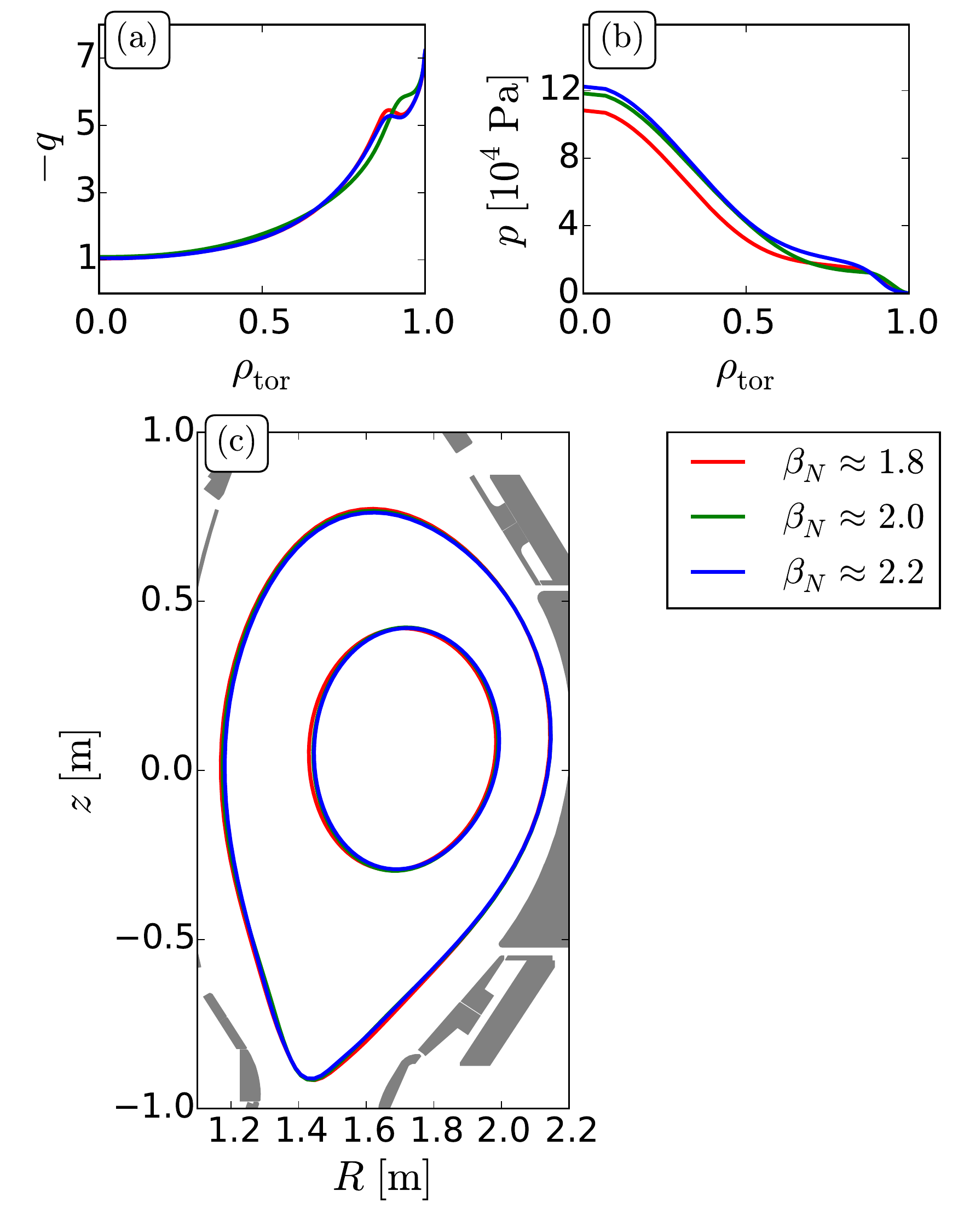} \\
 \caption{Properties of the three input equilibria. (a) q-profile and (b) pressure profiles versus \ac{rhotor}. (c)  $\ac{rhotor}=0.5$ and $1.0$ surface for low (red), medium (green) and high \ac{betaN} (blue) case. The corresponding shots are colored in table~\ref{tab:discharges}.}
 \label{fig:q-p-profile}
 \end{figure}

To account for changes in the $q$-profile and/or pressure profile, we use the 2D CLISTE equilibrium reconstructions from three different discharges to generate the input equilibria for the modeling of the displacement.
Figure \ref{fig:q-p-profile} shows the (a) $q$-profile, (b) pressure profile and (c) the shape of the \ac{LCFS} of the low (red), medium (green) and high (blue) \ac{betaN} case colored in table ~\ref{tab:discharges}. 

To avoid any influence of the rigid rotation and of \ac{ELM} physics on the equilibrium reconstruction, we use for the initial 2D equilibrium only pre-\ac{ELM} time-points from the magnetic signals  averaged over one rotation period. This is very similar to the procedure used for the synchronization of the \ac{ELM} cycle~\cite{Dunne:2012}.
Then, only pre-\ac{ELM} measurements from interferometry, \ac{LIB}, \ac{ECE}, \ac{CXRS} and  \ac{TS} are used for the edge pressure profile to further constrain the 2D equilibria (figure \ref{fig:q-p-profile}(a)). The \ac{ne} and \ac{Te} profiles from \ac{TS} are used to align the profiles. To account for the change in the edge pressure gradient due to the density pump-out, we use input equilibria once with strongest density 'pump-out' resulting in low \ac{betaN} (red) and once with almost no 'pump-out' leading to relatively high \ac{betaN} (blue). Since VMEC cannot handle the \ac{SOL}, we excluded  \ac{SOL} currents in the CLISTE equilibrium reconstruction. Moreover, the edge $q$ profile in CLISTE is constrained to match the Sauter predictions for the bootstrap current~\cite{Sauter:1999}  using \ac{ne} and \ac{Te} profile measurements mentioned previously (figure \ref{fig:q-p-profile}(a)).    
Consequently, the resulting edge $q$ and pressure profiles in the equilibrium have  experimental uncertainties depending on the measurements accuracy and on the alignment between the density and the temperature profiles. These uncertainties are taken into account by including one input equilibrium at medium \ac{betaN}, which exhibits a more outwardly shifted density profile~\cite{Dunne:2017}. This changes the edge pressure profile and the $q$-profile (green in figure \ref{fig:q-p-profile}). In total, we estimated that the uncertainties of the edge $q$-profiles are around $\pm10\%$.
The reason for this outward shift in the density profile is not clear. Since it is also present during the phases without \ac{MP}-field, we can rule out possible 3D effects as the cause. One possible explanation might be some changes in the divertor condition due to different wall conditions after the boronization, which is difficult to diagnose. 
However, the inclusion of the various  equilibria allows us to estimate the impact of such profile uncertainties in the input equilibria on the 3D distortion of the flux surfaces.
Although there are some variations in the equilibrium profiles in all cases, the  shape of the \ac{LCFS} is almost the same (figure \ref{fig:q-p-profile}(c)).

\subsection{Displacement calculated using the vacuum field approximation}

 \begin{figure}[ht]
   \centering
 \includegraphics[width=1.0\textwidth]{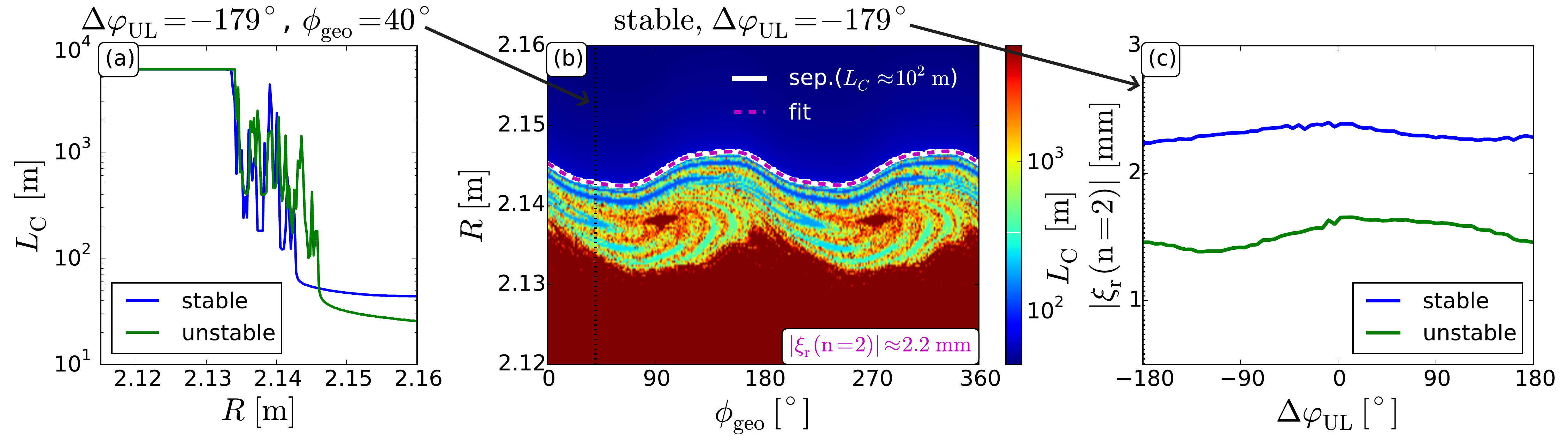} 
 \caption{(a) Connection length $L\rm{_C}$ along the \ac{LFS} midplane using the stable (blue) and the unstable (green) manifold at a single toroidal position $\phi{\rm{_{geo}}}$ and \ac{dphi}. (b) Laminar plot of $L\rm{_C}$ using the stable manifold along the toroidal coordinate $\phi{\rm{_{geo}}}$. $L\rm{_C}$ of $100\ \rm{m}$ to determine an 'effective' boundary and a displacement amplitude $\xi_r$. (c) The boundary displacement  using stable and unstable manifold versus \ac{dphi}. Amplitudes are not larger than $2.4\ \rm{mm}$.}
\label{fig:vacuumBoundary}
\end{figure}

To underline the importance of the plasma response,  measurements of the boundary displacement can be compared with predictions from vacuum field-line tracing. These predictions ignore shielding of the applied \ac{MP}-field and the amplification by stable ideal kink modes. The determination of a boundary displacement from vacuum field calculations is somewhat critical since a \ac{LCFS} is not necessarily preserved because of changes in the magnetic topology due to ergodization. However, one can estimate an 'effective' plasma boundary by a sudden increase of the connection length of the field-lines using equilibrium field superimposed with the applied \ac{MP}-field. Figure \ref{fig:vacuumBoundary}(a) shows the calculated \ac{LC} between the \ac{LOS} along the \ac{LFS} midplane and the target at one geometric toroidal coordinate $\phi\rm{_{geo}}=40^\circ$.
Since it is possible to follow the field-line in two directions, we refer to the direction towards the inner target as stable manifold and towards the outer target as unstable manifold (page 187 in Ref.~\cite{Abdullaev:2014}). The applied \ac{MP}-field  in figure~\ref{fig:vacuumBoundary}(a)  has $\ac{dphi}=-179^\circ$ ($\approx$vacuum \textit{resonant}). Field-line tracing is stopped when \ac{LC} reaches $6\ \rm{km}$. A sudden rise of \ac{LC} can be clearly identified at a \ac{LC} of around $10^2\ \rm{m}$, which is similar to the separatrix value in the axisymmetric case. Hence, we use this threshold ($\ac{LC}\approx10^2\ \rm{m}$) to define an 'effective' plasma boundary. These calculations are then extended to all toroidal positions, which allows us to determine an 'effective' $n=2$ boundary displacement. An example using the stable manifold is shown in figure \ref{fig:vacuumBoundary}(b). The solid white line is the 'effective' plasma boundary and the magenta dashed line is the sinusoidal fit to it. The derived amplitude is $2.2\ \rm{mm}$. This procedure is applied to all \acp{dphi} in the scan using both, the stable and the unstable manifolds (figure~\ref{fig:vacuumBoundary}(c)). 
The radial displacements for all cases are smaller than $\pm2.4\ \rm{mm}$. 

A combination of using the stable and the unstable manifolds does not increase the 'effective' boundary displacement. Instead, it leads to additional harmonic components. Additionally, the implementation of shielding on resonant surfaces would result in even smaller displacements.
For simplicity reasons, only the equilibrium with medium \ac{betaN} (green in figure~\ref{fig:q-p-profile}) is used for the vacuum field calculations. The choice of the equilibrium has only a small impact on the 'effective' boundary displacement evaluated using the vacuum field approximation.

\subsection{3D ideal MHD equilibrium calculations, VMEC}

In the experiments presented here \ac{Te} at the pedestal top is above $1\;\rm{keV}$ resulting in a low resistivity.  The perpendicular electron velocity is also high and has no zero-crossing in the pedestal (see also Section 4.3 in Ref.~\cite{Willensdorfer:2016}). Furthermore, 
we have no indication of mode penetration due to externally applied \acp{MP} like the \ac{Te} perturbations being in anti-phase on both sides of a rational surface or a  flattening of the profile at a rational surface. These reasons justify to compare the measured displacement with the one from an ideal \ac{MHD} code.

For the following comparison, we use a free boundary version of the ideal \ac{MHD} equilibrium code VMEC (also called NEMEC~\cite{Strumberger:2014}). 
VMEC is able to calculate the 3D distorted flux surfaces.  To parameterize its geometry it uses a Fourier representation.
VMEC minimizes the \ac{WMHD} by solving the variational problem $dW\rm{_{MHD}}/dt = 0$ using the steepest-descent moment method~\cite{Hirshman:1983}. 
To avoid time-consuming calculations which do not converge~\cite{Turnbull:2013}, we first test if the axisymmetric case (free boundary) converges using a small amount of flux surfaces ($\approx200$). If necessary, we adapt configuration parameters, such as the amount of poloidal mode numbers, and use the same  parameters for the extensive 3D cases~\cite{Wingen:2015}. These 3D calculations usually converge without any difficulties. 
To assure a sufficient resolution for all 3D calculations, we use 1001 flux surfaces, 17 toroidal mode numbers for one period ($\phi\rm{_{geo}}=0-180^\circ$ for the $n=2$ perturbation) including the negative ones ($n=-16,-14,\dots,14,16$) and 26 poloidal mode numbers. The choice of a sufficiently high resolved grid is essential, otherwise the calculated displacements can be underestimated (see \ref{appendix}). For this study, the input equilibria for all calculations are truncated at a normalized poloidal flux of 0.9999 (details about truncation in VMEC in Ref.~\cite{Wingen:2015}).

In total, we calculated 27 3D VMEC equilibria using three different input equilibria and nine different \ac{dphi} configurations. The main purpose of the variations in the $q$ and pressure profile of the input equilibria is to estimate their influence on the uncertainties of the displacement. 


\section{Plasma response as indicated by \ac{ELM} and density behavior}
\label{sec:ELMbehavior}

 \begin{figure*}[ht]
   \centering
 \includegraphics[width=0.9\textwidth]{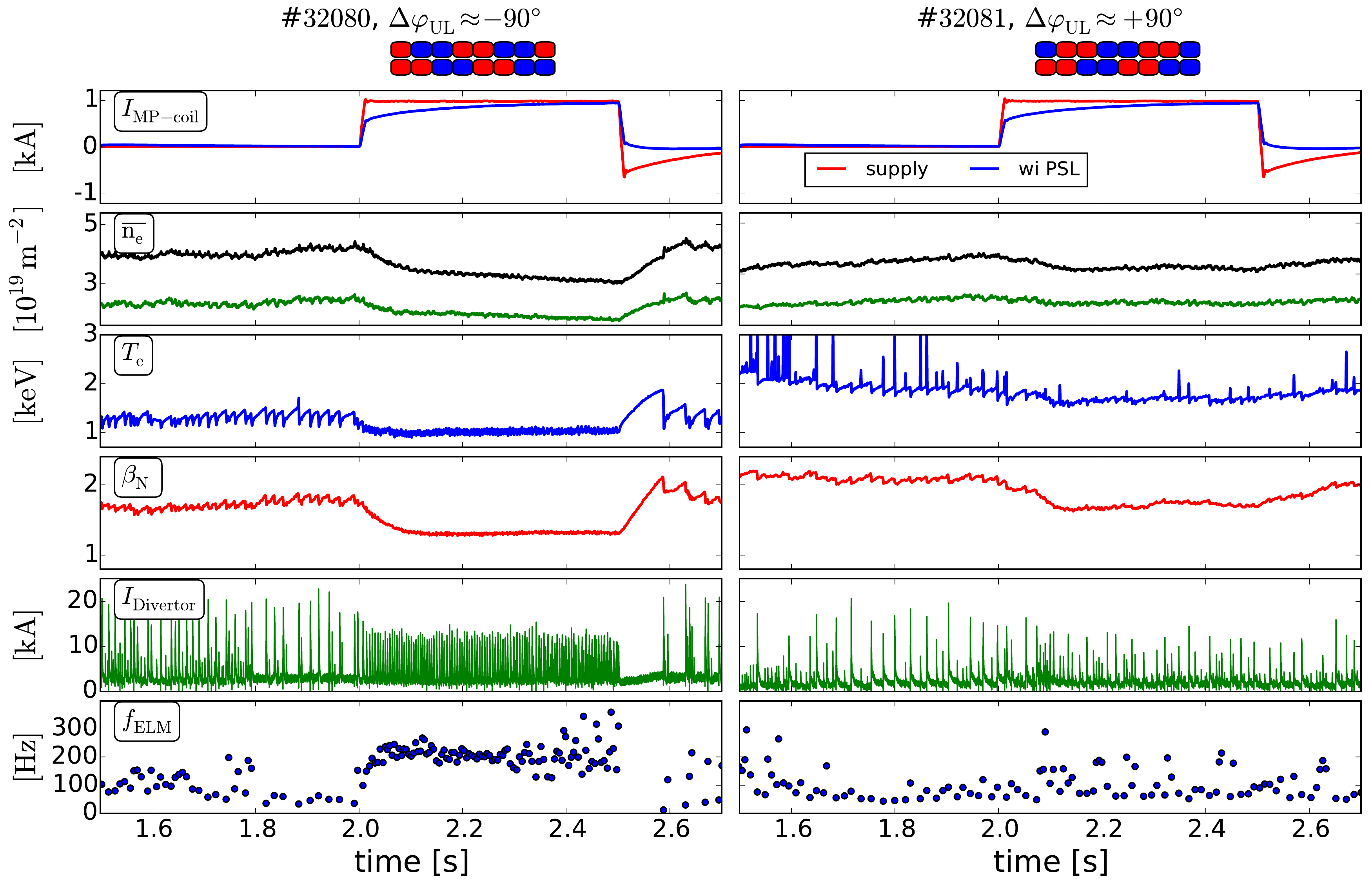} 
 \caption{Correlation between plasma response and ELM behavior. Time traces of discharge $\#32080$ using  \ac{dphi} of $-90^{\circ}$ (left) and discharge 32081 with  $+90^\circ$ (right). Insets at the top indicate the coil configurations. From top to bottom: \ac{MP}-coil supply current (red) and 'effective' (blue) current of one coil, core and edge chord of line integrated \ac{ne}, edge \ac{Te} from \ac{ECE}, normalized beta $\beta_N$, divertor current and \ac{ELM} frequency. \ac{ELM} behavior and \ac{ne}  are clearly changed for $\ac{dphi} \approx -90^{\circ}$.  }
\label{fig:TestScenario}
\end{figure*}

Empirically, the calculated displacement around the X-point, the top and the \ac{HFS} correlate with the \ac{ELM} frequency and the density 'pump-out' \cite{Kirk:2015,Suttrop:2017,Paz-Soldan:2015}. Hence, the simplest method to test the plasma response is to vary the applied mode spectrum via  \ac{dphi} and observe the change in \ac{ELM} frequency as well as density. This can be realized either by a continuous scan of \ac{dphi} or by applying several static \ac{MP}-phases with different \ac{dphi}. The first one has been done for a very similar plasma scenario, which only differs by a slight adaption of the upper shape to enable \acl{FILD} measurements~\cite{Garcia:2013}. This change is marginal suggesting a marginal impact on the plasma response~\cite{Li:2016}.
The comparison between MARS-F  calculations~\cite{Liu:2016} and the axisymmetric plasma response (e.g.~\ac{ne}, \ac{ELM} frequency) is shown in figure~7 in Ref.~\cite{Liu:2016b}. The strongest response in the measurements and calculations  are around $\ac{dphi}\approx-90^\circ$. The fact that $\ac{dphi}\approx-90^\circ$ is clear away from the optimum field alignment $\ac{dphi}\approx\pm180^\circ$ underlines the role of the stable ideal kink modes at poloidal mode numbers larger than the resonant components ($m>nq$).

 \begin{figure*}[ht]
   \centering
 \includegraphics[width=0.5\textwidth]{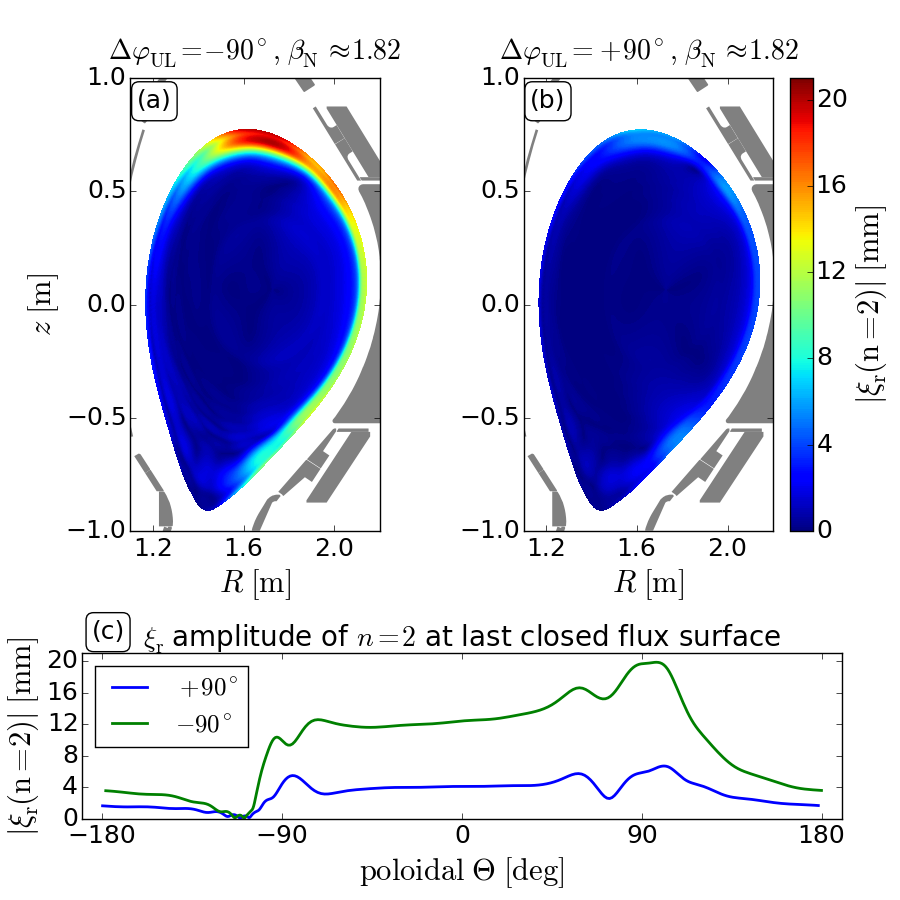} 
 \caption{ The poloidal distribution of the $n=2$ displacement amplitude applying (a) $\ac{dphi}\approx-90^{\circ}$ and (b) $\ac{dphi}\approx+90^{\circ}$ using the same input equilibrium with the same \ac{betaN}. (c) displacement amplitude at the \ac{LCFS} versus poloidal angle. The $-90^{\circ}$ case has larger displacements. }
\label{fig:displacmentTestScenario}
\end{figure*}

To verify this behavior for the identical configuration used for the rigid rotations, we conducted experiments applying static \ac{MP}-fields using $\ac{dphi}\approx-90^\circ$ (figure~\ref{fig:TestScenario} (left)) and $+90^\circ$ (right).
Clear changes in the confinement and \ac{ELM} frequency are observed depending on the applied coil configuration. To emphasize the effect of the \ac{MP}-field on $n_e$ and $T_e$, the \ac{MP}-field is switched-off 'fast' by compensating the image currents in the \ac{PSL} using counteracting coil currents \cite{Suttrop:2017,Leuthold:2017}. This compensation occurs within milliseconds and  is illustrated in the top frames of figure~\ref{fig:TestScenario} showing the applied coil current (red) of one \ac{MP}-coil and the corresponding effective coil current including the \ac{PSL} response (blue). During this 'fast' switch-off of the $-90^\circ$ configuration, \acp{ELM} disappear simultaneously with the \ac{MP}-field, which is typical for \ac{MHD} timescales (see Ref.~\cite{Suttrop:2017}). Afterwards, $n_e$ and $T_e$ recovers on transport time scales typical for the pedestal build-up after the transition from L- to \ac{H-mode} (see e.g.~Ref.~\cite{Willensdorfer:2014}). 
For $\ac{dphi}\approx+90^\circ$, almost no effect on the \ac{ELM} and density behavior is seen.

To analyze the role of the displacement, we calculated the corresponding  radial displacement using VMEC with the same \ac{betaN} as shown in figure~\ref{fig:displacmentTestScenario}. 
The $n=2$ displacement  amplitude, especially around the plasma top, is clearly stronger for the $-90^\circ$ (a) case than for  $+90^\circ$ (b). This emphasizes the effect of the displacement on the \ac{ELM} stability, particle and energy confinement.  Note, the \ac{LFS} and the \ac{HFS} responses of this plasma configuration have a very similar dependence on \ac{dphi}, which
is indicated in figure~\ref{fig:displacmentTestScenario}(c) and the following sections. This is a feature of the investigated plasma configuration and does not hold generally as suggested by calculations based on other  ASDEX Upgrade configurations~\cite{Liu:2016} and other machines~\cite{Paz-Soldan:2016}. 
One should also keep in mind that the displacement in the static experiments are expected to be roughly two times larger than the one in the rigid rotation experiments. This is because the static experiments have full current in each coil (see insets of figure~\ref{fig:TestScenario}), which is not possible in rotation experiments. This increases the field strength by a factor of $\sqrt{2}$. They also have no \ac{PSL} attenuation, which results in  a factor of $1.5$ stronger \ac{MP}-field. 
 
In summary, the importance of 3D \ac{MHD} physics on the \ac{ELM} stability and the particle transport is underlined by the effect of the different applied \acp{dphi} and by the timescales during the 'fast' switch-off. 

\section{Plasma top displacement}
\label{sec:plasmatop}

 \begin{figure}[ht]
 \centering
 \includegraphics[width=0.5\textwidth]{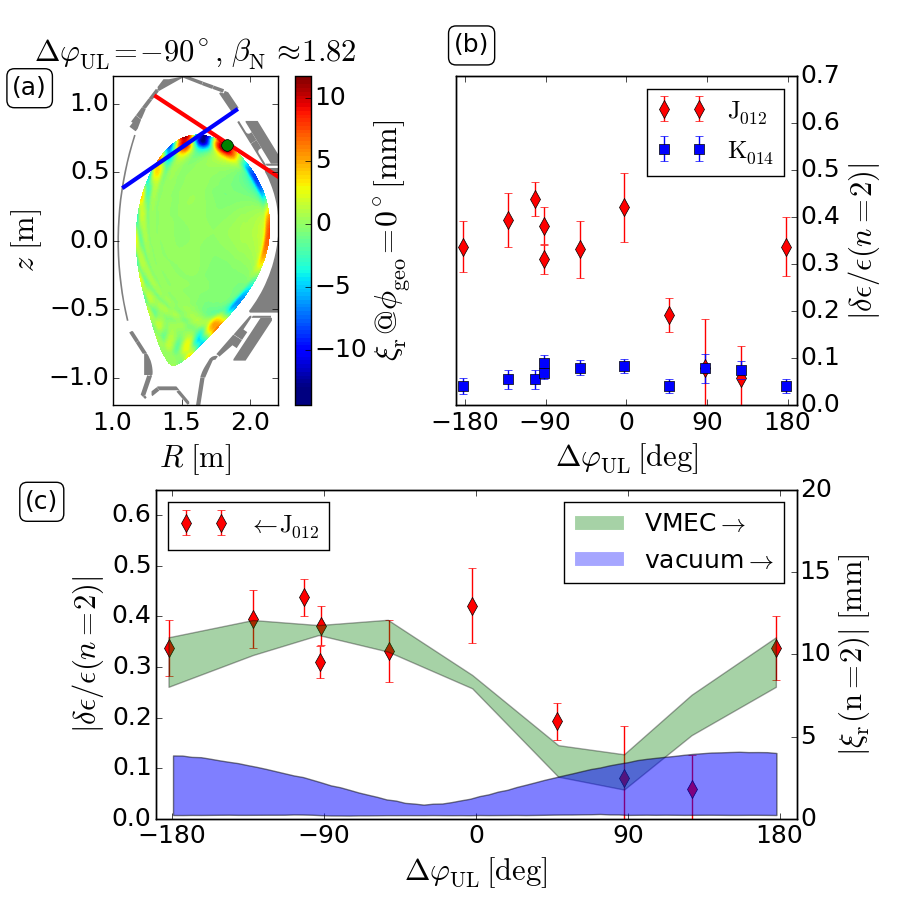} \\
 \caption{Comparing the displacements at the plasma top. (a) poloidal cut of the displacement $\xi\rm{_r}$ at one toroidal position $\phi\rm{_{geo}}=0^\circ$ and  \ac{LOS} of two soft X-ray channels. (b) relative emissivity ($\delta \epsilon/ \epsilon $) of $n=2$ versus \ac{dphi}  from the two channels in (a). (c) qualitative comparison relative emissivity from one channel with displacement amplitudes from VMEC and vacuum calculations at the position indicated by the green circle in (a).
 Only a qualitative comparison is possible. Displacements are largest around $\ac{dphi}\approx-90^{\circ}$.}
 \label{fig:SXRplasma}
 \end{figure} 
 
In this section, we further investigate the correlation between the applied poloidal mode spectrum $\ac{dphi}$ and  the displacement around the plasma top. 
We expect the strongest response at the plasma top around $\ac{dphi}\approx-90^\circ$ from the behavior of the \ac{ELM} frequency as well as the density 'pump-out' mentioned previously in section~\ref{sec:ELMbehavior}. 
This is clearly underlined by soft X-ray measurements viewing tangentially to the boundary of the plasma top (geometry in Fig~\ref{fig:SXRplasma}(a)). 

Figure \ref{fig:SXRplasma}(b) shows the relative emissivity ($\delta \epsilon/ \epsilon $)  of the $n=2$ perturbation from two soft X-ray channels determined from the various rigid rotation phase versus \ac{dphi}. One point corresponds to one rigid rotation phase and one channel. The relative emissivity from the channel $J_{012}$ clearly peaks around $\ac{dphi}\approx-90^\circ$.  This channel is almost perfectly tangential to the boundary and its relative emissivity is therefore a good indicator for the displacement. To illustrate that channels which are not  perfectly tangential to the boundary do not deliver useful displacement data, we add measurements from a second channel $K_{014}$. This  channel is not able to resolve the perturbation structures, because it simultaneously views the maximum and minimum displacement as demonstrated by the poloidal cut in figure \ref{fig:SXRplasma}(a). Thus, the  perturbation in the emissivity is always small and not a good measure for the displacement. 

However, the measurements from channel $J_{012}$ can be used to qualitatively compare them to the amplitude of radial displacement calculated  using VMEC and the vacuum field approximation shown in  Fig.~\ref{fig:SXRplasma}(c). The radial displacement is calculated where the channel \ac{LOS} crosses the boundary indicated by a green circle in Fig.~\ref{fig:SXRplasma}(a). 
The green shaded area in Fig.~\ref{fig:SXRplasma}(c) shows the possible VMEC solutions and the blue shaded are the possible solutions using the vacuum field approximation. For the comparison, only the amplitudes of the dominant toroidal component ($n=2$) are shown. 
The predicted VMEC and observed displacement amplitudes  have their maxima at  $\ac{dphi}\approx-90^\circ$ as well as minima at around $+90^\circ$ and correlate strongly with  the \ac{ELM} behaviour from the previous section.  
The vacuum field approximation does not reflect the \ac{dphi} dependency at all and predicts clearly lower displacements than VMEC.

\section{\ac{LFS} midplane displacement}
\label{sec:midplane}

 \begin{figure}[ht]
 \centering
 \includegraphics[width=0.6\textwidth]{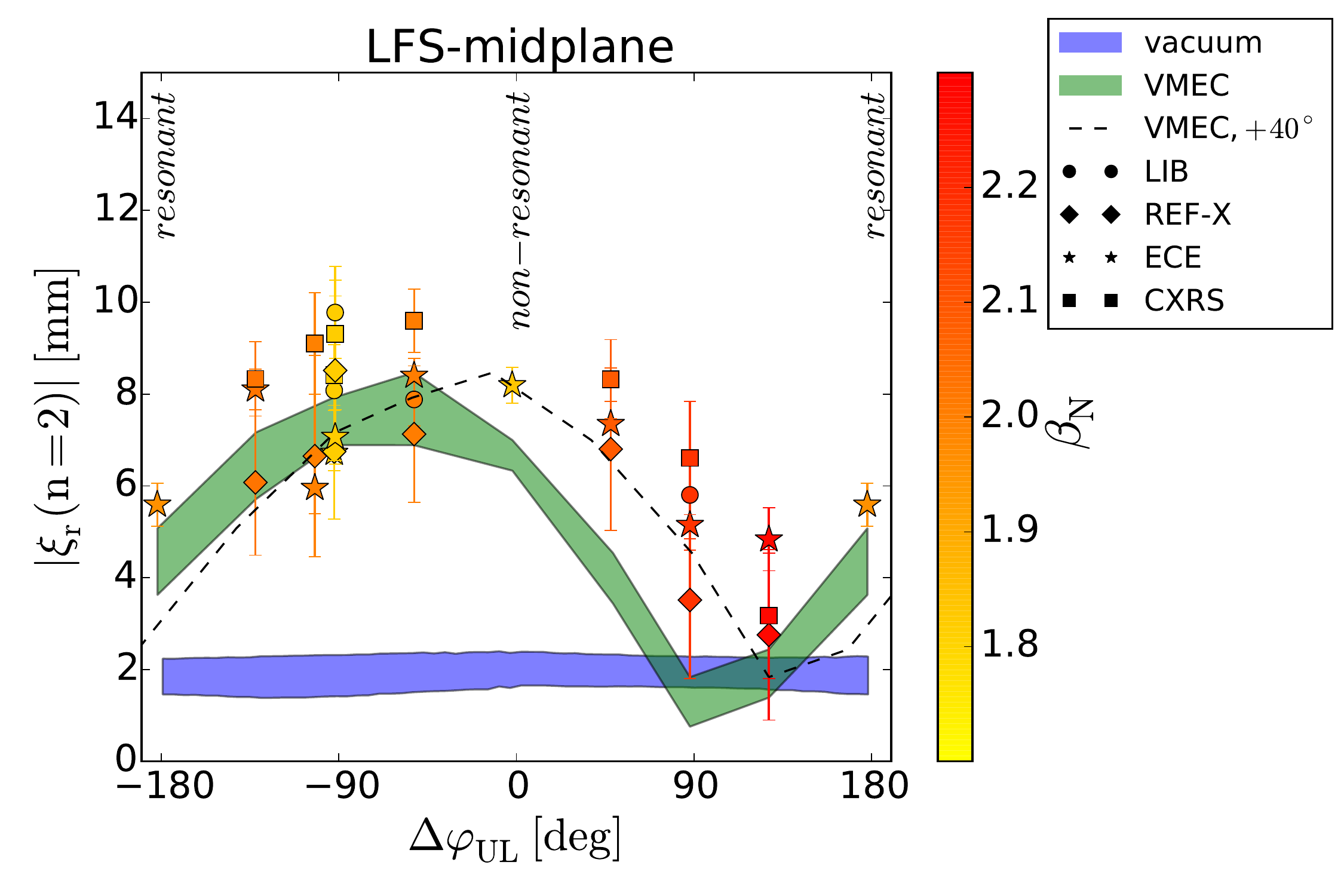} \\
 \caption{ Amplitude of the $n=2$ radial displacements versus \ac{dphi}. Each symbol is the measured displacement of one rigid rotation phase determined by one diagnostic. The color scaling indicates the measured \ac{betaN}. The possible solutions using the vacuum field approximation and VMEC are indicated by blue and green shaded areas, respectively. The dashed line shows the maxima of the VMEC solutions shifted by $40^\circ$. Except for a shift of around $40^\circ$, VMEC and the measurements agree well. }
 \label{fig:MidDisplacement}
 \end{figure} 

In the previous section, the radial displacement between the predictions from VMEC and the measurements were qualitatively compared. In this section, we go a step further and aim to make a quantitative comparison using the high resolution diagnostics around the \ac{LFS} midplane. To increase the accuracy of the analysis, the effects of the \ac{PCS} are included (section~\ref{sec:pcs}). Although the \ac{LFS} response is thought to play a minor role in the \ac{ELM}-mitigation~\cite{Paz-Soldan:2016}, this comparison is very valuable to benchmark VMEC.

To merge the various displacement measurements  around the \ac{LFS} midplane using  different diagnostics and experiments into one comparison, it is necessary to add the following considerations to the analysis: (i) there is one rigid rotation using $3\ \rm{Hz}$. To compare it to the $2\ \rm{Hz}$ experiments, we simply multiply the evaluated displacement at $3\ \rm{Hz}$ by 1.08  to account for the additional attenuation due to the \ac{PSL} response. This factor comes from the ratio between the $2\ \rm{Hz}$ and $3\ \rm{Hz}$ attenuation  (see section~\ref{sec:PSLresponse}). (ii) The various \ac{LOS} of the profile diagnostics are not exactly perpendicular to the axisymmetric surface. Hence, we map the displacement onto the normal using the axisymmetric shape from figure~\ref{fig:q-p-profile}, which allows us to compare it with the calculated radial displacement. The largest impact on the displacement amplitude is seen in the case of the \ac{LIB} geometry, where it changes by only $5\%$. (iii) The diagnostics are not exactly located on the \ac{LFS} midplane. To account for poloidal asymmetries, we scale the measured displacement to the midplane using the ratio between the calculated displacement at the \ac{LOS} and the midplane. To get this ratio, we use the average of the three VMEC calculations at the corresponding \ac{dphi}. Again, the evaluations of the \ac{LIB} measurements are primarily affected and the 'worst' case requires a change of only $18\%$, which is less than $1\ \rm{mm}$.

Figure~\ref{fig:MidDisplacement} shows the radial displacement amplitudes ($n=2$) around the \ac{LFS} midplane versus \ac{dphi} from the measurements, from the VMEC solutions (green shaded area) as well as from the vacuum field calculations (blue shaded areas).
Good agreement can be found between the measurements and the VMEC calculations. When the stable ideal kink modes are expected to be excited, both clearly surpass  the prediction from the vacuum field calculations, no matter which manifold is used. 
It is also seen from the range of the green shaded area that the observed variation in \ac{betaN}, in the edge pressure gradient and in the $q$-profile have no large impact on the calculated displacement amplitude from VMEC.  Additionally,  a slightly different choice of the used resolution in VMEC can increase the calculated displacement amplitude by $1\;\rm{mm}$ (see \ref{appendix}). Then, the agreement with the measurements would be even better.
We also observe no systematic difference in the displacement between density and temperature measurements, which underlines the presence of perturbed flux surfaces.


An offset of $40^\circ$ in \ac{dphi} between the measurements and the VMEC calculations indicates a minor disagreement. This offset is outside the measurement uncertainties of the displacement amplitude and outside the possible range of VMEC calculations.
The applied poloidal mode spectra and thus, the \ac{dphi} dependence of the plasma response depends strongly on  the positions of the flux surfaces due to the Grad-Shafranov shift, thus, \ac{betaN} and on the $q$-profile.
The considered variations in the $q$-profile (figure~\ref{fig:q-p-profile}) alone shifts the \ac{dphi} dependence of the resonant components from vacuum field calculations by $30^\circ$. But  as seen in figure~\ref{fig:MidDisplacement}, the \ac{LFS} response is only shifted by $10^\circ$. Thus, the considered variations in the $q$- and pressure profile are not enough to explain these discrepancies in \ac{dphi} in the case of the \ac{LFS} response. It should be noted that the  $q$- and pressure profile have not been varied independently.



\section{Summary and Discussion}
\label{sec:conclusion}

The main goal of this paper was to quantitatively compare measurements of the boundary displacement to ideal \ac{MHD} modeling using VMEC. Rigidly rotating $n=2$ \ac{MP}-fields with different \ac{dphi} and toroidally localized diagnostics deliver the accuracy which is needed to  measure  the displacement and its dependence on the applied poloidal mode spectrum.
To keep the margins for interpretation from the experimental side small, we included various profile diagnostics in the analysis. Furthermore, we accounted for additional plasma movements due to the plasma control system and varied the applied poloidal mode spectrum using \ac{dphi}. For both the experimental analysis and the modeling input, only pre-\ac{ELM} data were used.
On the modeling side, we consider the \ac{MP}-field attenuation due to passive conductors (\ac{PSL}),  the changes in the edge pressure profile due to the density 'pump-out' and  the small variations in the $q$-profile.
To avoid any misinterpretation due to 'inadequate' grid settings, we also performed sensitivity studies on the grid resolution (see \ref{appendix}). 

From this comprehensive study, we conclude that VMEC correctly predicts the boundary displacement amplitude due to stable ideal kink modes excited by external \ac{MP}-fields. Good quantitative  agreement around the \ac{LFS} midplane is found. The \ac{HFS} response around the plasma top could only be compared qualitatively  mainly due to the lack of locally available diagnostics.
Although VMEC cannot resolve localized sheet currents, assumes nested flux surface and hence, has no resistive \ac{MHD}, no \ac{SOL} physics and no toroidal rotation included~\cite{Wingen:2017}, it reproduces the amplitude of the displacement and its dependences on \ac{dphi}. The only caveat is that there is a systematic offset of around $40^\circ$ in \ac{dphi} between the measurements and the modeling around the \ac{LFS} midplane. This motivates further studies on the impact of the $q$-profile on the \ac{dphi} behavior. To rule out a lack of physics like \ac{SOL} currents, two-fluid \ac{MHD}~\cite{Ferraro:2013}, toroidal rotation~\cite{Wingen:2017} or localized sheet currents~\cite{Loizu:2016} as a possible explanation for the shift in \ac{dphi}, further comparisons to other MHD codes \cite{King:2015, Reiman:2015} like IPEC, JOREK, MARS-F and M3D-C1 are needed.

In conclusion, we can state that, if no strong resistive \ac{MHD} mode activity like mode penetration is present, VMEC can properly compute the 3D perturbation of the flux surface at the boundary.

\section{Acknowledgement}

M.W. would like to thank J.~Loizu for fruitful discussions.
This work has been carried out within the framework of the EUROfusion Consortium and has received funding from the Euratom research and training programme 2014-2018 under grant agreement No 633053. The views and opinions expressed herein do not necessarily reflect those of the European Commission.

\appendix

\section{Grid resolution of VMEC}
\label{appendix}
The grid resolution  in VMEC is often discussed~\cite{Turnbull:2012, Wingen:2015,Lazerson:2016} and is crucial for a reliable prediction from VMEC. To avoid any misinterpretation because of a too low grid resolution, we scanned the number of flux surface, of poloidal and of toroidal mode numbers. The default setting for this study is 1001 flux surfaces, 17 toroidal mode numbers and 26 poloidal mode numbers for one $n=2$ period ($\phi \rm{_{geo}}=0-180^\circ$).

\subsection{Number of flux surfaces}
 \begin{figure}[ht]
 \centering
 \includegraphics[width=0.5\textwidth]{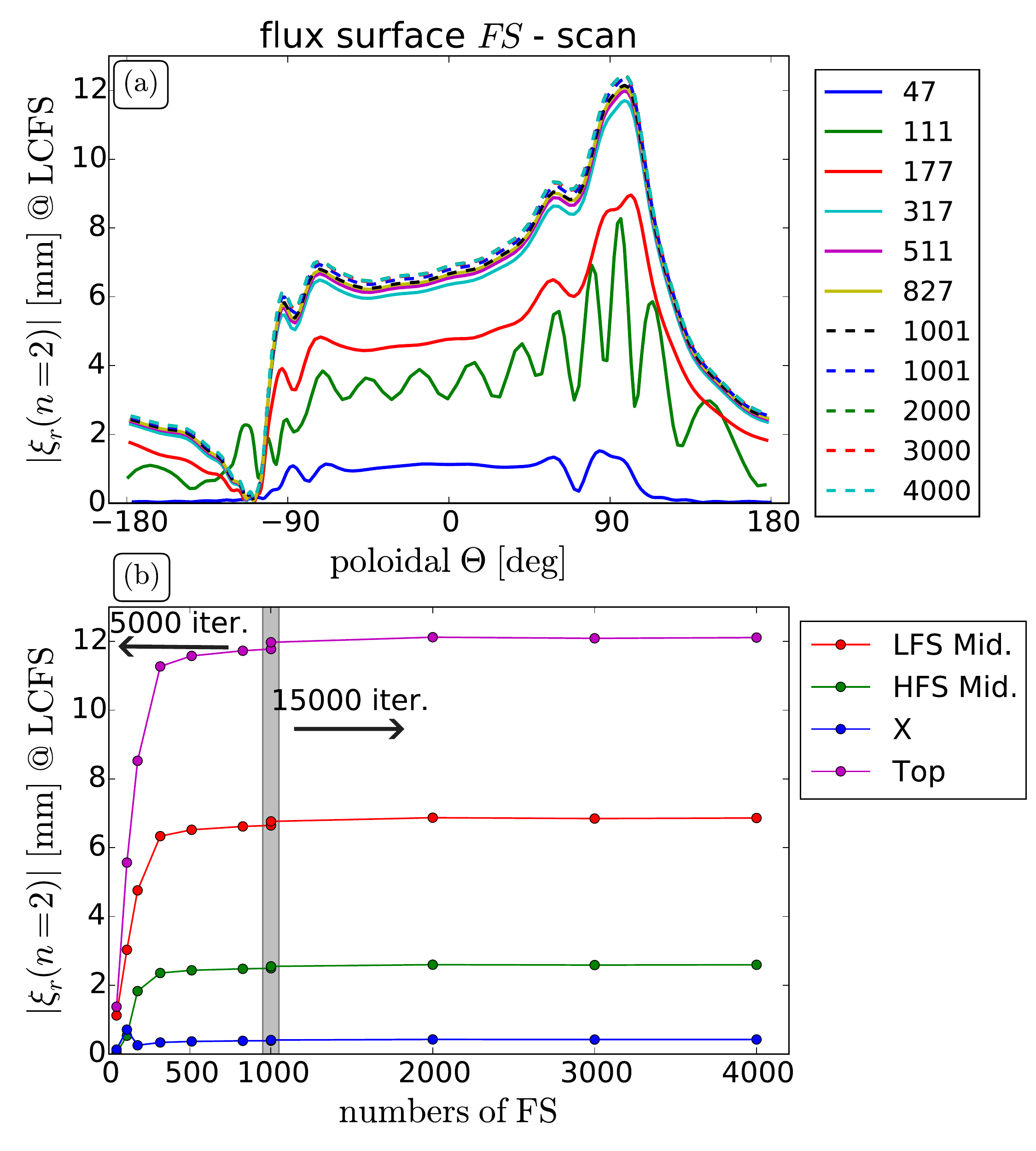} \\
 \caption{Sensitivity study on the numbers of flux surfaces (FS). (a) The $n=2$ displacement amplitude at the \ac{LCFS} versus poloidal angle $\Theta$. (b) Displacement versus numbers of FS for specific poloidal positions. The grey bar indicates the default resolution.}
 \label{fig:FSscan}
 \end{figure} 

To study the impact of the radial resolution on the resulting displacement~\cite{Lazerson:2016}, we increased the amount of flux surfaces up to 4000. We pick a case with a strong plasma response ($\ac{dphi}\approx-90^{\circ}$, low \ac{betaN} and $2\;\rm{Hz}$). Figure \ref{fig:FSscan} shows the sensitivity study on the number of flux surfaces. The displacement at the \ac{LCFS} versus the poloidal angle is shown in figure \ref{fig:FSscan}(a). Figure \ref{fig:FSscan}(b) illustrates the displacement at the \ac{LFS} midplane ($\Theta=0^\circ$), plasma top ($+90^\circ$),    \ac{LFS} midplane  ($\pm180^\circ$) and the X-point. One should note that for less than 1000 flux surfaces, 5000 iterations are used, whereas for more than 1000 surfaces 15000 iterations are used. However even with 4000 flux surfaces, the maximum displacement does not change more than $0.2\;\rm{mm}$ with respect to the used number of 1001. This is not a surprise, since (i) VMEC uses an equidistant toroidal flux grid (in this version), which is relatively dense towards the edge and (ii) we are studying an experimental configuration which primarily exhibits edge perturbations. 

\subsection{Number of poloidal mode numbers}

 \begin{figure}[ht]
 \centering
 \includegraphics[width=0.5\textwidth]{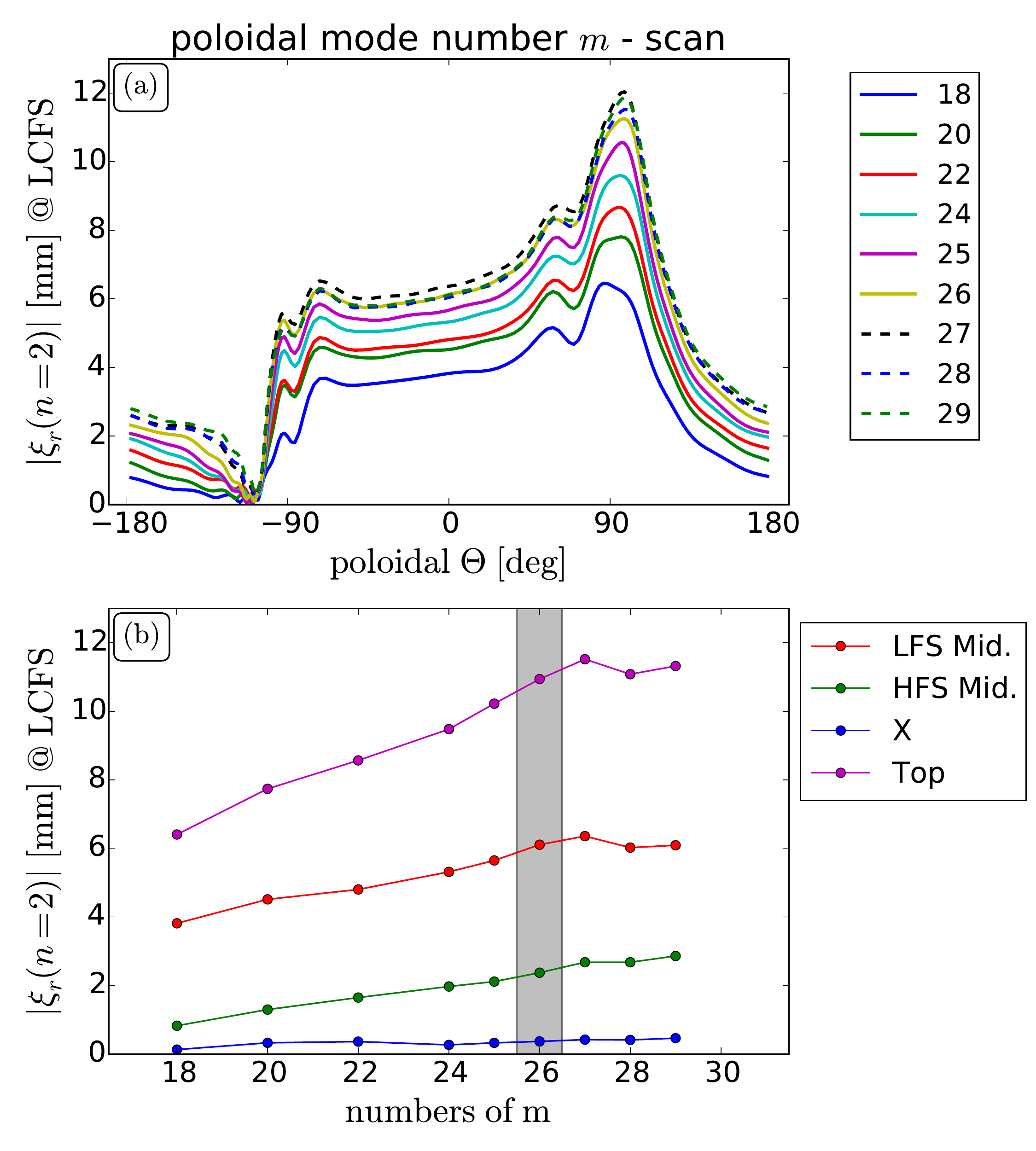} \\
 \caption{Sensitivity study on the numbers of poloidal mode numbers $m$. (a) The $n=2$ displacement amplitude at the \ac{LCFS} versus poloidal angle $\Theta$. (b) Displacement versus numbers of $m$ for specific poloidal positions. The grey bar indicates the default resolution.}
 \label{fig:mscan}
 \end{figure}
In this and the following comparison, we use the $3\;\rm{Hz}$ \ac{MP}-field attenuation .
A minimum of 18 poloidal mode numbers is required to reproduce the axisymmetric elongated shape. This number is still far too low to get reasonable displacement  values, because they increase until they stagnate around 26 (Figure~\ref{fig:mscan}(b)).
  
 \subsection{Number of toroidal mode numbers}

  \begin{figure}[ht]
 \centering
 \includegraphics[width=0.5\textwidth]{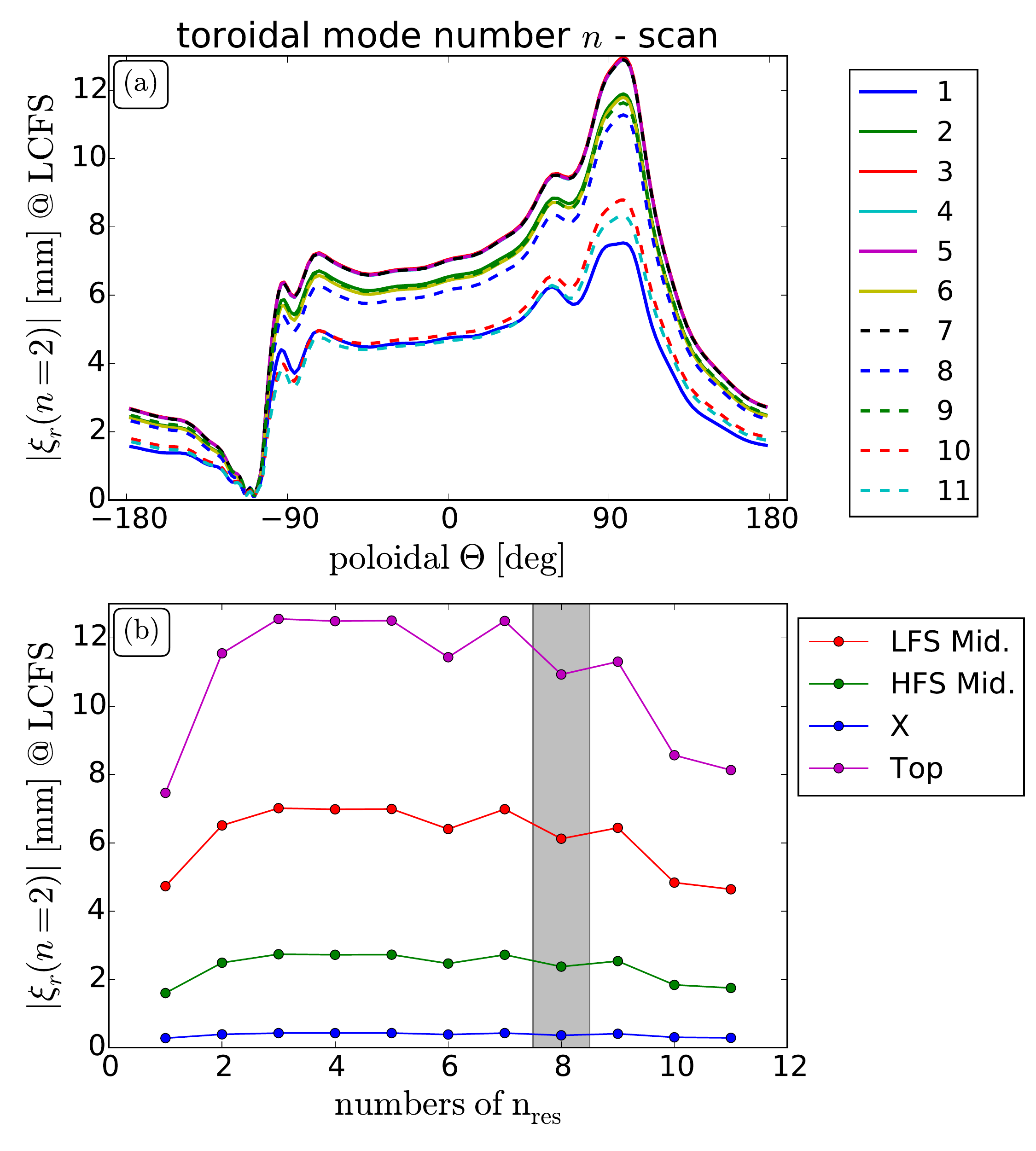} \\
 \caption{Sensitivity study on the numbers of toroidal mode numbers $n{_{res}}$ using one period ($\phi\rm{_{geo}}=0-180^\circ$). (a) The $n=2$ displacement amplitude at the \ac{LCFS} versus poloidal angle $\Theta$. (b) Displacement versus numbers of $n{_{res}}$ for specific poloidal positions. The grey bar indicates the default resolution. $n{_{res}}=8$ means that we consider up to $n=16$.}
 \label{fig:nscan}
 \end{figure} 

The input parameter in  VMEC for the toroidal resolution  $n{_{res}}$ also accounts for the negative mode number. Employing  $n{_{res}}=8$ for one $n=2$ period ($\phi \rm{_{geo}}=0-180^\circ$) means 17 toroidal mode numbers. The \ac{zeta} are used to describe the vacuum field perturbations from the \ac{MP}-coils for the boundary condition. Usually, we set \ac{zeta} four time larger than $n{_{res}}$. During this sensitivity study, it turned out that, at least,  $\ac{zeta}=32$ are required to describe the $n=2$ vacuum field perturbations for one period (4 coils in each row). Otherwise the VMEC calculations did not converge. So for $n{_{res}}<9$, we used a \ac{zeta} of 32 and otherwise, four times of $n{_{res}}$.
Figure \ref{fig:nscan} shows the sensitivity study. Already $n{_{res}}=3$ delivers reasonable results. But using $n{_{res}}=3$ instead of $n{_{res}}=8$ does not save a lot of computational time, since the vacuum calculations with  $\ac{zeta}=32$ are the most time consuming part. However, for larger toroidal mode numbers $n{_{res}}>8$ the amplitude of the dominant mode decreases. A similar behavior is given, when too many harmonics of a sine fine are given to fit experimental data. Then,  the amplitude of $n=2$  decreases with increasing harmonics as well. We assume that this is also the case when the amount of toroidal mode numbers in VMEC increases. 



\vspace{10mm}

\bibliographystyle{unsrt}
\bibliography{displacement}

\end{document}